\documentclass[aps,prb,floats,epsf,twocolumn,showpacs]{revtex4}
 \usepackage{graphics, times, wrapfig}
\usepackage{slashed}

\begin{document}

\title{Multi-critical behavior of $Z_2\times O(2)$ Gross-Neveu-Yukawa theory in graphene}

\author{Bitan Roy}

\affiliation{Department of Physics, Simon Fraser University, Burnaby, British Columbia, V5A 1S6, Canada}

\begin{abstract}
Multi-critical behavior of interacting fermions in graphene's honeycomb lattice is presented. In particular, we considered the spin triplet insulating orders, where the spin rotational symmetry of the order parameter is explicitly broken. By casting the problem in terms of Gross-Neveu-Yukawa theory we show that such symmetry-breaking terms are irrelevant near the metal-insulator critical point. A finite Yukawa coupling among bosons and fermions improves the stability of such critical point against the symmetry-breaking perturbations. Physical sources of such symmetry-breaking terms are pointed out. Critical exponents are calculated near the transitions as well.
\end{abstract}


\maketitle

\vspace{10pt}

The pseudo relativistic nature of low energy quasi particles opened a new frontier in condensed matter physics with the discovery of graphene, a monolayer of graphite.\cite{geim} Even though in its pristine state graphene appears to be a semi-metal, fermions may also find themselves in various insulating phases as well. Depending on the relative strength of the different finite range components of the Coulomb interaction, the ground state can lack a plethora of discreet or continuous symmetries. For instance, a sufficiently strong onsite Hubbard interaction (U) orients the spin at each site in an opposite direction from its neighboring ones, whereas the average electron density  acquires a staggered pattern at large nearest-neighbor Coulomb repulsion $(V_1)$.\cite{semenoff,igor1,bitan} Yet another insulating phase may result from a strong next nearest neighbor Coulomb repulsion, which can induce a gapped insulating phase with finite circulating currents between the sites on the same sublattice.\cite{Haldane,raghu} This state is named the \textit{quantum anomalous Hall} (QAH) state and breaks time reversal symmetry (TRS) upon acquiring a finite expectation value in the ordered state. However, fluctuations preempt appearance of a QAH state and stabilize the \textit{quamtum spin Hall} (QSH) insulator. The QSH state breaks the TRS only for each spin component.\cite{kane,bernevig} These transitions out of the symmetric semi metallic phase into the gapped insulating phases are believed to be continuous and belong to the Gross-Neveu universality class.\cite{igor1,bitan,oskar} Recently it has been argued that Kekule bond density wave order may also appear when the nearest neighbor and second nearest neighbor interactions are comparable.\cite{franz} If the net interaction has an attractive component, fermions may find various superconducting states to condense into. For example, an onsite attraction favors a spin singlet s-wave superconducting ground state \cite{arun}. A second nearest neighbor attractive interaction leads to a spin triplet f-wave superconductor.\cite{honercamp} A spatially inhomogeneous superconducting ground state, which breaks the translational symmetry of the honeycomb lattice into Kekule pattern of bond order parameter is proposed as a variational ground state for strong attractive interactions between the electrons living on nearest neighbor sites. \cite{royherb, uchoa}

In recent studies \cite{bitan,oskar}, it has been argued that near the semimetal-insulator quantum critical points the pseudo relativistic invariance of the non-interacting theory is restored. Consequently, the dynamical critical exponent (z) is exactly equal to \textit{unity} leading to a non-critical behavior of Fermi velocity. There the nature of the quantum phase transitions in both spin singlet and triplet channels is considered, assuming that the interacting Hamiltonian describing the transitions does not break the spin rotational symmetry. However, the symmetry can be broken by, for example, the crystal strain or finite spin-orbit coupling.\cite{kane} Once we introduce an anisotropy in the order parameter along one spin direction, theory loses the $O(3)$ symmetry associated with the spin rotation, and enjoys a reduced $Z_2 \times O(2)$ symmetry. In the present discussion, I am concerned with the relevance of the broken spin rotational symmetry of the order parameters near the quantum criticality. Within the framework of the one loop $\epsilon-$ expansion, I found that such anisotropy in the spin degrees of freedom is irrelevant in the vicinity of the metal-insulator quantum critical point. However in a similar situation in an interacting bosonic system, it has been shown that the order-disorder transition is driven by a \textit{biconal} fixed point, enjoying the $Z_2 \times O(2)$ symmetry.\cite{calabrese} In this discussion, I will show that the critical point associated with the disorder-order transition exhibits full rotational symmetry. However, this may be a consequence of the one loop calculations. An interesting observation is that in presence of finite Yukawa coupling among the Dirac fermions and the self interacting Higgs fields, such anisotropy acquires an additional degree of irrelevance near the transitions. Here the Higgs bosons are composite fields.

Let us first consider a collection of free fermions on graphene's honeycomb lattice. The tight binding model for spin-1/2 fermions, on graphene honeycomb lattice with only nearest-neighbor hopping, is defined as
\begin{equation}
H_{t}=-t \sum_{\vec{A},i, \sigma=\pm 1} u_{\sigma}^{\dagger}(\vec{A}) v_{\sigma}(\vec{A} + \vec{b}_i) +H.c.,
\end{equation}
where $u_\sigma,v_\sigma$ are fermioninc operators on two triangular sub-lattices of the honeycmb lattice, and $\vec{b}_i$s, with $i=1,2,3$ connect each sites on A-sublattice with its three nearest-neighbor ones. Keeping the Fourier modes near two inequivalent Dirac points, located at the corners of the Brillouin zone \cite{semenoff}, at $\vec{K}=(1,1/\sqrt{3}) 2\pi/a \sqrt{3}$ and $\vec{K}'=-\vec{K}$, let us construct an $8-$ component Dirac fermion $\Psi=(\Psi_+,\Psi_-)^{\top}$, with
\begin{equation}
\Psi^\top _{\sigma}  (\vec{q}) =
\left[ u _\sigma (\vec{K} + \vec{q}), v _\sigma (\vec{K}+ \vec{q}), u_\sigma (-\vec{K}+ \vec{q}), v_\sigma (-\vec{K}+ \vec{q}) \right],
\end{equation}
with $\sigma=\pm$, which corresponds to electrons spin projection along the z-axis. The tight-binding Hamiltonian in the low energy approximation then takes the form 
\begin{equation}
H_t=\sum_{\vec{q}} \Psi^{\dagger}(\vec{q}) H_D \Psi (\vec{q}) +O(q^2),
\end{equation}
with $H_D$ being the Dirac Hamiltonian in two dimensions which in our representation acquires a simple form
\begin{equation}
H_D=\sigma_0 \otimes i \gamma_0 \gamma_i q_i.
\end{equation} 
Here the four-component anticommuting gamma matrices belong to the representation $\gamma_0=\tau_0\otimes\tau_3,\gamma_1=\tau_3\otimes\tau_2,\gamma_2=\tau_0\otimes\tau_1$, where $\tau_0$ is the two componenet identity matrix and $\vec{\tau}$ are the standard Pauli matrices.\cite{bitan} We also define the remaining two gamma matrices $\gamma_3=\tau_1\otimes\tau_2,\gamma_5=\tau_2\otimes \tau_2$. The two component Pauli matrices $(\sigma_0,\vec{\sigma})$ act on the spin indices. Here, for convenience we set the Fermi velocity $v_F=ta \sqrt{3}/2$ to unity. Before we consider the effect of electron-electron interaction on the quasi-particle dispersion, it is useful to register the symmetries of the free Hamiltonian. The Dirac Hamiltonian commutes with $P=\sigma_0\otimes i \gamma_{3}\gamma_5$, which in our representation stands for the generator of translation. It also commutes with $I_K=\sigma_0 \otimes i \gamma_1 \gamma_5$ and $I_{uv}=\sigma_0 \otimes \gamma_2$, upon inverting the momentum axes $q_1 \rightarrow -q_1$ and $q_2 \rightarrow -q_2$, respectively. These two operators, respectively correspond to the exchange of two non-equivalent Dirac points and the sub-lattices.

We now consider the effect of electron-electron interaction on the gapless excitations spectrum in the vicinity of the Dirac points. Due to vanishing density of states at Fermi energy, the gapless excitations are robust against any small electron-electron interaction. However, if the interactions are sufficiently strong, the system may suffer semi-metal insulator transitions. In $2+1$ dimensions there are two ways of generating a dynamical mass at sufficiently strong interactions. The first order-parameter (OP) is
\begin{equation}
\phi=(\phi_s,\vec{\phi}_t)=(\langle \Psi^\dagger \sigma_0 \otimes \gamma_0 \Psi \rangle,\langle \Psi^\dagger \vec{\sigma}\otimes \gamma_0 \Psi \rangle).
\end{equation}
$\phi_s (\vec{\phi}_t)$ preserves (breaks) the TRS and breaks chiral $U_c(4)$ symmetry (CS) generated by $\{\sigma_0,\vec{\sigma}\}\otimes \{I_4,\gamma_3,\gamma_5,\gamma_{35}\}$, where $\gamma_{35}= i \gamma_3 \gamma_5$. The TRS is defined as
$\Psi_{\sigma} \rightarrow I_t \Psi_{\sigma}$, where $I_t$ is an anti-unitary operator defined as $I_t=U K$, $K$ being the complex conjugate. Here $U$ is the unitary part and in `graphene representation' $U=i\gamma_1 \gamma_5$. In our representation $\phi_s$ corresponds to a finite chemical potential differing in its sign from its neighbors, whereas $\vec{\phi_t}$ to a finite N$\acute{e}$el ordering. The second order parameter
\begin{equation}
\chi=(\chi_s,\vec{\chi}_t)=(\langle \Psi^\dagger \sigma_0 \otimes i \gamma_1 \gamma_2 \Psi \rangle,
\langle \Psi^\dagger \vec{\sigma} \otimes i \gamma_1 \gamma_2 \Psi \rangle),
\end{equation}
on the other hand breaks the TRS for each spin component but preserves the CS. Such a correlated ground state supports an intra sub-lattice current, introduced by Haldane \cite{Haldane}, circulating in opposite directions on two sub-lattices.

It is worth mentioning that the singlet components of the OPs $(\chi_s,\phi_s)$ break the Ising like symmetry among the sub-lattices, whereas the triplet OPs $(\vec{\chi_t},\vec{\phi_t})$, additionally break the full spin rotational symmetry. Therefore the quantum critical points (QCPs) corresponding to the development of $\chi_s$ and $\phi_t$ belong to the same universality class. Analogously the QCPs associated with the generation of $\vec{\chi_t}$ and $\vec{\phi_t}$  also enjoy the same universality class.\cite{oskar} Moreover all the OPs listed above transform as scalars under the pseudo Lorentz transformation. Therefore near the quantum critical points, driving the system out of the semi metallic phase to the ordered insulating phases the \textit{dynamical critical exponent(z)} is expected to be \textit{unity}. Nevertheless, a Loretz symmetry breaking perturbation is found to be irrelevant near the QCPs.\cite{bitan,oskar}

Next we concentrate on the spin triplet insulating orders, namely $(\vec{\chi_t},\vec{\phi_t})$ and at the same time set $\chi_s=\phi_s=0$. Let us start with the onsite Hubbard interaction (U) only, for simplicity. However, we introduce a non-trivial anisotropy along one particular spin projection. On methodological level, it appears interesting to study the relevance of such anisotropy near the semimetal-insulator quantum critical point. The quantum mechanical action in the presence of anisotropic interaction along one particular spin projection reads as $S=\int^{1/T}_0 d\tau d\vec{x} L_{int}$, where 
\begin{equation}
L_{int}=\bar{\Psi} \sigma_0 \otimes \gamma_{\mu} \partial_{\mu} \Psi + g_{\parallel}(\Psi^\dagger \vec{\sigma_{\parallel}} \otimes \gamma_0 \Psi)^2
+g_{\perp}  (\Psi^\dagger \sigma_3 \otimes \gamma_0 \Psi)^2,
\end{equation} 
where $\vec{\sigma_\parallel}=(\sigma_1,\sigma_2)$ and $\mu=0,1,2$. $\mu=0$ stands for the imaginary time component. The Einstein summation convention is assumed, but only over the repeated space time indices. Within the framework of the Hubbard model with only onsite repulsion one discovers $g_\parallel=g_\perp=U/16$.\cite{herbut2}

By performing the Hubbard-Stratonovich transformation one can rewrite the effective action corresponding to $L_{int}$ in $d-$ dimension as
\begin{eqnarray}\label{Action_aniso}
&S&=\int d^{d}x \{ -[ \bar{\Psi} \sigma_0 \otimes \slashed{\partial} \Psi 
+ {g}_{\parallel}\vec{\phi_{\parallel}}(\Psi^\dagger \vec{\sigma_{\parallel}} \otimes \gamma_0 \Psi) \nonumber \\
&+&g_{\perp}\phi_{\perp} (\Psi^\dagger \sigma_3\otimes \gamma_0 \Psi)]
+ \frac{1}{2} [(\partial_{\mu} \vec{\phi_{\parallel}})^2 + (\partial_{\mu} \phi_{\perp})^2]  
+ m_{\parallel} \vec{\phi}^2_{\parallel}\nonumber \\
&+& m_{\perp} \vec{\phi_{\perp}}^2 
+ \lambda_{\parallel} (\sum_{i=1,2} \phi^2_{i})^{2} + \lambda_{\perp} \phi^{4}_{3} 
+ \frac {\widetilde{\lambda}}{12} (\sum_{i=1,2} {\phi_i}^2 {\phi_3}^2)  \}. \nonumber \\
\end{eqnarray}
Here the fermion bilinears are coupled to the Higgs OPs, $\phi_i$s. In the Yukawa form the theory is renormalizable in $3+1$ dimension, where both Yukawa $(g^2_{\parallel},g^2_{\perp})$ and Higgs self energy couplings $(\lambda_{\parallel},\lambda_{\perp},\widetilde{\lambda})$ are essentially dimensionless. In particular, when $g^2_{\parallel}=g^2_{\perp}$, concomitantly 
$\lambda_{\parallel}=\lambda_{\perp}=\widetilde{\lambda}$, $\vec{\phi}$ corresponds to N$\acute{e}$el order parameter.

Next, we consider the renormalization group study of the coupling constants in the Yukawa theory, Eq.\ [\ref{Action_aniso}]. Here we restrict ourselves only to the one loop expansion. In particular, we are interested in studying its flow equations in $d=4-\epsilon$ dimensions. The coupling constants $(\lambda_\perp, \lambda_\parallel, \widetilde{\lambda},g_\perp,g_\parallel)$ are dimensionless in $d=4$, and hence a controlled perturbative expansion can be performed in terms of a small parameter $\epsilon$, define above \cite{Justin}. For similar studies in a system of interacting bosons, readers may consult Ref.\cite{bookHerbut}. Subscribing to the fact that the pseudo relativistic invariance is respected near the QCPs \cite{bitan,oskar}, upon integrating out the \emph{fast} bosonic and fermionic modes within the four-momentum shell $\Lambda/b < (\omega^2+\vec{k}^2)^{1/2} < \Lambda$ $(b>1)$, one can write down the differential flow equations of the coupling constants as

\begin{equation}\label{para}
\frac{d g^2_{\parallel}}{d \ln{b}}= \epsilon g^2_{\parallel}- (2N+2)g^4_{\parallel}+ g^2_{\parallel}g^2_{\perp},
\end{equation}

\begin{equation}\label{perp}
\frac{d g^2_{\perp}}{d \ln{b}}= \epsilon g^2_{\perp}- (2N+3)g^4_{\perp}+ 2 g^2_{\parallel}g^2_{\perp},
\end{equation}

\begin{equation}
\frac{d \lambda_{\parallel}}{d \ln{b}} = \epsilon \lambda_{\parallel}- (\frac{5}{3}\lambda^2_{\parallel}+\frac{1}{6}\widetilde{\lambda}^2)
- 4 N \lambda_{\parallel} g^2_{\parallel}  + 24 N g^4_{\parallel} ,
\end{equation}

\begin{equation}
\frac{d \lambda_{\perp}}{d \ln{b}}= \epsilon  \lambda_{\perp}- (\frac{3}{2}\lambda^2_{\perp}+\frac{1}{3}\widetilde{\lambda}^2)
- 4 N \lambda_{\perp} g^2_{\perp} + 24 N g^4_{\perp},
\end{equation}

\begin{eqnarray}
\frac{d \widetilde{\lambda}}{d \ln{b}}&=&\epsilon \widetilde{\lambda}- (\widetilde{\lambda}\lambda_{\parallel}
+\frac{1}{6}\lambda_{\parallel} \lambda_{\perp}+\frac{2}{3}\widetilde{\lambda}^2) 
- 2 N \lambda_{\perp} (g^2_{\perp}+g^2_{\parallel}) \nonumber \\
&+& 24 N g^2_{\parallel} g^2_{\perp},
\end{eqnarray}
where $g_x={S_d} \Lambda^{\epsilon} g_x, \quad \lambda_x={S_d} \Lambda^{\epsilon} \lambda_x, \quad x=\parallel,\perp$, and $\widetilde{\lambda}=S_d \Lambda^{\epsilon} \widetilde{\lambda}$. $S_d$ is the surface of a $d-$ dimensional unit sphere and $N$ corresponds to the number of $4-$ component spinor. Thus for graphene $N$ is equal to $2$. Here the ultraviolet cut-off, $\Lambda \approx 1/a$, corresponds to the interval of energy over which the linear approximation of the density of states holds.\cite{Paiva} 
\begin{figure}[t]
{\centering\resizebox*{40mm}{!}{\includegraphics{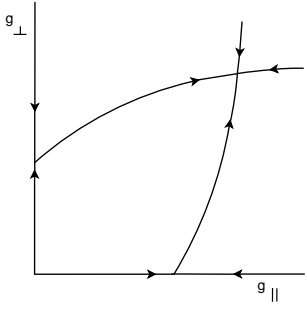}}
\par} \caption[] {Flow diagram in the Yukawa coupling space. Fully stable fixed point is located at $g^2_{\parallel}=g^2_{\perp}=\frac{\epsilon}{2 N+1}= \frac{\epsilon}{5}$ for $N=2$.}\label{Yukawa}
\end{figure}
First we study the flow of the Yukawa couplings in the $(g^2_{\parallel},g^2_{\perp})$ plane. Since we consider the flow of the coupling constants in the critical hyperplane, we wish to find the fixed point which is stable from all directions. From the first two flow equations, Eq.\ [\ref{para},\ref{perp}], one finds that the fixed point in this plane stable from all direction is located at
\begin{equation}
g^2_{\parallel}=g^2_{\perp}=\frac{\epsilon}{2 N+1}.
\end{equation}
Therefore near the quantum critical point, two Yukawa couplings, $(g^2_{\parallel},g^2_{\perp})$ enjoy equal strength, Fig.\ \ref{Yukawa}. This feature is independent of the value of $N$. Hence for the rest of our discussion we set $g^2_{\parallel}=g^2_{\perp}$. Next we consider the Higgs sector of the theory. Upon setting both the Yukawa couplings equal ($=\epsilon/5$ for $N=2$), we found the fixed point, in the $(\lambda_{\parallel},\lambda_{\perp},\widetilde{\lambda})$ plane stable from all direction, resides at
\begin{equation}
\lambda_{\parallel}=\lambda_{\perp}=\widetilde{\lambda}=\frac{48}{55} \epsilon.
\end{equation}
Therefore only the fixed point with all three Higgs and both the Yukawa couplings having equal strength has one unstable direction and thus is \textit{ critical}.\cite{critplane} The unstable direction corresponds to flow of the mass of the Higgs field, $m_{\parallel}=m_{\perp}=m$ and its flow reads as
\begin{equation}\label{mass}
\frac{d m^2}{d \ln{b}}=2 m^2-\frac{5}{6} \lambda m^2-2 N g^2 m^2,
\end{equation} 
where $g^2_{\parallel}=g^2_{\perp}=g$ and $\lambda_{\parallel}=\lambda_{\perp}=\widetilde{\lambda}=\lambda$. The mass of the Higgs field (m) is proportional to the temperature. Therefore, near the metal-insulator quantum critical point anisotropic coupling turned out to be irrelevant and consequently the spin rotational symmetry is restored. The emergence of $SO(3)$ symmetry near the QCP happens to take place for arbitrary N. A renormalization group study on a closely related, but purely bosonic $\Phi^4$ theory $(g_\parallel=g_\perp =0)$, is performed over a most general $O(n_1)\oplus O(n_2)$ symmetric Landau-Ginzburg-Wilson Hamiltonian, involving two fields $\phi_1$ and $\phi_2$ enriched by $n_1$ and $n_2$ components, respectively. In an extensive five loop $\epsilon-$ expansion it was found that for $n_1=1$ and $n_2=2$, the disorder-order transition is governed by a \textit{biconal} fixed point possessing a $Z_2\times O(2)$ symmetry. However, the critical exponents near the \textit{biconal} fixed point are extremely close to the ones  for the $SO(3)$ symmetric \textit{Heisenberg} fixed point.\cite{calabrese} It is worth mentioning that the critical point associated with the order-disorder transition enjoys the $O(2)$ symmetry even though the interactions are invariant under a $Z_2\times Z_2$ symmetries. On the other hand, when $n_1+n_2 > 3$, the system finds itself in an ordered phase via \textit{decoupled} fixed points. It is admitted that the emergence of the spin rotational symmetry near the quantum critical point may be an artifact of the one loop calculations. However, it is worth mentioning that the negative eigenvalues of the stability matrix near the spin symmetric fixed point announced above, acquires additional contributions of the same sign for finite Yukawa couplings. Therefore, the flow of the irrelevant trajectories toward the critical point is \textit{faster} when the Higgs fields are coupled to the fermions. By taking $g_{\parallel}=g_\perp=0$ the eigenvalues of the stability matrix at $\lambda_{\parallel}=\lambda_{\perp}=\widetilde{\lambda}=\frac{48}{55} \epsilon$ are $\left(-1,-7/11,-2/11 \right) \epsilon$, whereas those with $g_{\parallel}=g_\perp=\epsilon/5$ are $\left(-19/5,-177/55,-137/55 \right) \epsilon$, with $N=2$. Therefore it may be worth studying the multi-critical behavior of the present model beyond one loop level.

Finally let us consider the other triplet order $\vec{\chi}_t= \langle \Psi^\dagger \vec{\sigma} \otimes i \gamma_1 \gamma_2 \Psi \rangle$, that breaks the TRS for each spin component. The insulating state is referred as \textit{quantum spin Hall insulator} (QSHI). Upon incorporating the fluctuations in the ordered phase around the saddle point, it also preempts appearance of \textit{quantum anomalous Hall insulator}. A sufficiently strong second nearest neighbor repulsion can take the system into such an ordered phase.\cite{Haldane,raghu} The spin rotational symmetry of this order parameter is broken in the presence of a finite spin-orbit coupling, which is proportional to the third component of the order parameter.\cite{kane} In a graphene system the spin-orbit coupling $\sim 0.01-0.2 K$, is extremely small in comparison to the finite range Coulomb repulsions.\cite{min-yao} Even though we cannot study the critical properties of the semimetal-insulator instability toward the QSHI phase, at $d=2+1$ it belongs to the Gross-Neveu universality class of the anti-ferromagnetic instability. This is because the term $\bar{\Psi} [\vec{\sigma}\otimes i \gamma_3 \gamma_5] \Psi$ can be transformed into $\bar{\Psi} [\vec{\sigma}\otimes I_4] \Psi$, while keeping the free Lagrangian invariant.\cite{bitan} Therefore, in the presence of a finite spin-orbit coupling the order parameter $\vec{\chi}_t$ lives on the surface of the $S_2$, sphere in three dimensions.

To summarize, we here considered the spin triplet insulating order of Dirac fermions, when the spin rotational symmetry of the order parameter is explicitly broken. Performing the Hubbard-Stratonovic transformation, we present the theory as a $Z_2 \times O(2)$ symmetric Gross-Neveu-Yukawa theory. Within the framework of one loop $\epsilon-$ expansion, where $\epsilon=4-d$, we found a quantum critical point that restores the symmetry under spin rotation and drives the system from a symmetric semimetallic phase to a ordered insulating phase.\cite{oskar} Even though the conclusion is based on a simple one loop calculation, we found that such QCP acquires extra stability when the bosonic Higgs fields are coupled to fermionic fields via Yukawa interactions. This result dictates that even when there exists anisotropy in an interacting model at the lattice scale, along different spin directions, it smears out near the transition and restores the symmetry under spin rotation. Recent quantum Hall experiments \cite{ong}, revealed a Kosterlitz-Thouless scaling of the resistivity near the metal insulator transition, when the system is tuned to filling one-half. This observation initiated a search for the origin of the vortex like excitations of the Dirac fermions. One candidate for the possible order parameter with the requisite $U(1)$ structure is the Kekule bond-density-wave, which is favored by the electron-phonon coupling.\cite{Nomura, Hou} Another possibility is the antiferromagnetic order \cite{igorQHE,Mcdonald}, favored by onsite Hubbard interaction, projected onto an easy plane by Zeeman coupling. In the presence of a magnetic field, the relativistically invariant band, collapses onto set of Landau levels (LL) and existence of a half-filled LL drives the system through metal-insulator transition even at infinitesimal interactions. At zero field criticality gap shows a perfect square root dependence on the magnetic field. If the Hubbard interaction is not too far from its critical (zero field) strength for insulation \cite{Paiva, katsnelson}, scaling (sublinear) of the interaction induced gap is essentially determined by the zero field critical properties. The magnetic field then plays the role of finite size length scale. Such scaling function has been computed field theoretically, as well as numerically in the Hartree limit\cite{igorQHE, roy-catalysis}. In this work, I showed that the Ne$\acute{e}$l order is unlikely to be restricted in the easy plane, within the frame work of the anisotropic Hubbard model. It is admitted that the source of anisotropy in the Hubbard model is currently unknown. However, a finite Zeeman coupling immediately restricts the N$\acute{e}$el order in the easy plane and may support vortex excitations.\cite{herbut2,royherb}

Finally one can compute the critical exponents near the critical point which restores the spin rotational symmetry. These have been also previously computed in Ref. \cite{oskar}. Namely, the correlation length exponent can be calculated from Eq.\ [\ref{mass}], yielding $\nu=\frac{1}{2}+\frac{21}{55} \epsilon$. Moreover one can compute the bosonic anomalous dimension, which is $\eta_{b} = \frac{4}{5} \epsilon$, and the fermionic anomalous dimension $\eta_{\psi}=\frac{3}{10} \epsilon$ as well. The fermionic anomalous dimension determines the behavior of the fermion propagator $G^{-1}_{\psi} \sim (\omega^2+ k^2)^{(1-\eta_{\psi})/2}$, near the critical point as one approaches the QCP from the semimetallic side of the transition.

The author is thankful to V. Juri\v ci\' c and C. Chamon for valuable discussions. Author acknowledges a great debt to I.F. Herbut for useful discussions and the critical reading of the manuscript. This work was supported by NSERC of Canada. Kelly Cheng and Payam Mousavi are also gratefully thanked for some useful comments on the manuscript.

\end{document}